\documentclass{article}
\pdfoutput=1

\usepackage[preprint, nonatbib]{nips_2018}

\usepackage[utf8]{inputenc} % allow utf-8 input
\usepackage[T1]{fontenc}    % use 8-bit T1 fonts
\usepackage{url, hyperref}  % hyperlinks
\usepackage{booktabs}       % professional-quality tables
\usepackage{amsfonts}       % blackboard math symbols
\usepackage{nicefrac}       % compact symbols for 1/2, etc.
\usepackage{microtype}      % microtypography
\usepackage{enumerate}
\usepackage{multirow}
\usepackage[pdftex]{graphicx}

\title{In (Stochastic) Search of a Fairer Alife}

\author{
  Dmitriy Volinskiy\thanks{Dmitriy Volinskiy and Lana Cuthbertson are Data Scientist and Director of Customer Experience Strategy, 
  respectively, with ATB Financial; {\tt https://ai.atb.com}.} \\
  ATB Financial\\
  Edmonton, AB, Canada \\
  \texttt{dvolinskiy@atb.com} \\
   \And
   Lana Cuthbertson \\
   ATB Financial \\
   Edmonton, AB, Canada \\
   \texttt{lcuthbertson@atb.com} \\
   \And
   Omid Ardakanian\thanks{Omid Ardakanian is with the Department of Computing Science, University of Alberta, Canada; {\tt http://webdocs.cs.ualberta.ca/$\sim$oardakan/}.} \\
   University of Alberta \\
   Edmonton, AB, Canada \\
   \texttt{ardakanian@ualberta.ca} \\
}

\begin{document}

\maketitle

\begin{abstract}
Economies and societal structures in general are complex stochastic systems which may not lend themselves well to algebraic analysis. An addition of subjective value criteria to the mechanics of interacting agents will further complicate analysis. The purpose of this short study is to demonstrate capabilities of agent-based computational economics to be a platform for fairness or equity analysis in both a broad and practical sense.
\end{abstract}

%==================================
\section{Introduction}\label{sec:intro}
%==================================

Introducing subjective values such as fairness 
to formal fields of inquiry has been a challenge, particularly so in economics. 
Much to the chagrin of positive economics proponents, 
there is a large and still rapidly growing body of literature on fairness, 
distributive equity and other seemingly not-so-positive analysis. 
Taxation, health or welfare economics are most commonly the fields where the concept of equity or economic equality has most support. 
Things have been trickier in microeconomic analysis, risk, and game theory.
Most utility-theoretic risk and non-cooperative game models assume self-interest only.
The pioneering behavioral economics research of the 80s and the 90s (e.g.,~\cite{Berg95,Kahneman86,Rabin93}) 
provided much broader insight into problems that had traditionally 
been up an economist's alley. 
Cooperative reciprocity was introduced, utility and value functions were revised 
to add a social component. 
Yet fairness still has a long way to go to become a mainstream consideration 
in neoclassical economics.

Fairness of course is not confined to games, utility and general equilibrium analyses. 
There are direct parallels between definitions from various fields of science. 
For example, fairness is well-studied in computer systems and communication networks,
mainly in the context of resource allocation;
there exist several axiomatically justified fairness criteria
which differ in the choice of the objective function.
\emph{Max-min fairness} is achieved by allocating the available resources 
in a particular feasible way that an attempt to allocate more 
to an agent will result in a decrease in the allocation of some other agent 
featuring an equal or smaller allocation~\cite{Kelly98}.
Thus, it is a form of Pareto efficiency with a value statement added to it.
\emph{Proportional fairness} maximizes the sum of the log-utility functions of all users~\cite{Kelly98}.
It is a scale invariant Pareto optimal solution,
which is consistent with axioms of fairness in game theory.
Equilibrium solutions to cooperative games as applied to the flow control problem
in computer networks~\cite{Mazumdar91} 
have their direct counterparts in economic risk and game-theoretic models.
In recent years, fairness has become a central concern in machine learning 
as it is being used to address key societal challenges~\cite{barocas18}.

Arguably, where most formal studies of fairness 
need a tad of improvement is practical application. 
Economics or not, an equation-based study typically looks into properties 
or a particular fairness-inducing decision criterion: 
a value function, an allocative mechanism, a twist to a classical game, or similar. 
But more general questions that a policy-maker or a business leader 
may ask often remain unanswered. 
Would a society of a certain fair design grow and prosper? 
Would adopting or eliminating a certain business practice help us preempt future regulation? 
Enter the realm of Agent-based Computational Economics (ACE). 
ACE is the computational study of economic processes modeled as dynamic systems of interacting agents, an alife\,---\,an artificial life form instances~\cite{tesfatsion06}. 
The agents do not need to be literally individuals or businesses\,---\,an agent is a behavioral rule or a set of rules, 
so agents can be biological entities, physical entities, 
hierarchical structures, or even nodes in a decision process. 
Rich and flexible an interpretive platform as it is, ACE has not yet been largely adopted for studies of social system features, 
although some inroads have been made on this front~\cite{terano2003meeting,Heckbert10}.

The purpose of this short study is to demonstrate ACE capabilities 
to be a platform for fairness or equity analysis in both broad and practical sense. 
We show how to set up a simple alife society 
with some fairness features to be investigated, 
and then demonstrate how to use the mechanics of evolutionary computation\,---\,an evolution-inspired trial-and-error type of stochastic search\,---\,to get some practical insight into what living in such a society would look like. 
And this comes with some unexpected findings.

%==================================
\section{Experiment Setup}\label{sec:formulation}
%==================================
Consider a society of $N$ agents, indexed by $i$, each of which maximizes their (Cobb-Douglas) utility 
function given by
\begin{equation}
U_i(z_i,c_i) = z_i^{\alpha_i} c_i^{\beta_i}, 
\end{equation}
where $z_i$ denotes leisure time, $c_i$ denotes consumption, and $l_i$ denotes labor input. 
Hence, we have $z_i+l_i=24$ and $c_i=l_i(l_i+\sum_{-i}l_j)^{\gamma-1}$ ($\gamma>1$),
where the parameters are preferences of the individual, $(\alpha_i,\beta_i)$, 
and the productivity of the society’s transformation function, $\gamma$. 
Note that every individual in this setup receives a consumption share of the society's output, $(l_i+\sum_{-i}l_j)^{\gamma}$ which is proportional to their contribution $l_i$. 
We further assume homogeneity of degree one for everyone's utility, i.e., $\alpha_i+\beta_i=1, \forall i$; we will provide a rationale for this assumption later in this section.

If each individual maximizes their utility in a myopic fashion, or equivalently, assumes input of others to the economy,
$\sum_{-i}l_j$, to be exogenous (and very large compared to their own $l_i$), the setup in (1) is trivial and has a simple algebraic solution. 
Now let us introduce some complications. First, we assume that each individual maximizes 
a $\sigma_i$-weighted sum of their own utility and utility of their children, 
with an integer $k_i\geq0$ becoming a parameter of choice as to how many kids to have. 
The utility criterion in turn becomes:
\begin{equation}
U_i^* = U_i\left(\frac{z_i}{k_i+1},\frac{c_i}{k_i+1}\right) + \sigma_i(k_i)U_i(z_i,c_i).
\end{equation}
Notice the trade-off: increasing the desired number of kids increases one's utility through the second term\,---\,the agent effectively assumes that, utility-wise, their children are going to be just like them, 
``like father like son''\,---\,whilst decreasing their utility though the first term: the parent can no longer enjoy as much leisure nor can they consume at the original level. 
Second, we introduce time as a dimension so that $U_i^*$ becomes $U_{it}^*$, as well as a mortality function 
$m_{it}=m\left(\sum_{s=0}^t l_{is}\right)$, 
where $m(.)$ is some strictly monotone function with values in the range $(0,1)$. 
Thus, the more the agent works over their life to a certain point, the sooner they are going to die.

Even if we assume that an agent optimizes while being oblivious to their mortality (which otherwise would become an intertemporal choice problem), algebra quickly gets murky here. 
Since we have assumed that $\alpha_i+\beta_i=1$, i.e. if one doubles the inputs, it doubles the utility output, 
one can observe that an agent will choose to have at least one kid if $\sigma_i>0.5$, 
as the 50\% loss in their ``own'' utility will be compensated through the utility of their offspring, and so on. 
Last but not least, it is obvious that, to optimize the economy's output, 
a central planner will be needed to choose all $l_i$ as opposed to agents selecting those individually. ACE time?

%==================================
\section{Simulation: Strategies, Process, Metrics}\label{sec:simulation}
%==================================
We define four strategies shown in Table~\ref{tbl:strategies} for the purpose of our simulation exercise.
The process then proceeds as a sequence of steps forming an evolutionary algorithm (e.g. \cite{fogel1998evolutionary}), 
for each of these strategies.
\begin{table}[ht]
  \caption{Simulation Strategies.}
  \label{tbl:strategies}
  \centering
  \begin{tabular}{l|p{11cm}}
    \toprule
    \textbf{Strategy 0} & Each agent maximizes their own utility assuming the myopic contribution to the society's production function. This can be shown to be equivalent to maximizing the sum or average utility, i.e., that of a representative agent. 
    This is the least ``socialist'' strategy and the baseline.\\ \midrule
    \textbf{Strategy A} & We optimize the utility of a representative agent; however, the production gets optimized globally, by a ``central planner'' 
    who determines everyone's contribution to the society's production function.\\ \midrule
    \textbf{Strategy b} & Each agent assumes the myopic contribution to the society's production function; however, we choose to maximize the minimum utility in the society.\\ \midrule
    \textbf{Strategy Ab}& We maximize the minimum utility in the society. The production gets optimized by choosing everyone's input by the central planner. 
    This is the most ``socialist'' strategy. \\
    \bottomrule
  \end{tabular}
\end{table}

One hundred societies are created for a single experiment run, each lasting 100 generations 
in case a society proves to be a successful one.
\begin{enumerate}[i.]
    \item Generate a population of agents using the standard Uniform as the proposal distribution for the parameters $(\alpha_i,\beta_i,\sigma_i)$, and a draw from an Exponential distribution plus one for $\gamma$.
    \item Choose the optimal values for $l_i$ and $k_i$ for the chosen strategy using the Nelder-Mead simplex method~\cite{olsson1975nelder} (one needs to use a gradient-free optimization technique as $k_i$ needs to be an integer).
    \item Allow agents to mate: we do not introduce any sexual dimorphism in the model 
    and let the agents mate based on the similarity of their preference parameters.
    \item Each formed pair produces their offspring in a quantity set by a draw 
    from a Poisson distribution with the mean equal to the average of $k_i$ and $k_j$; 
    preference parameters of the offspring are weighted averages 
    of those of their parents with a small random mutation component added.
    \item A portion of the parents population is set to die off by drawing 
    from a Bernoulli distribution with the probability if the agent’s demise 
    equal to the value of their $m_{it}$.
    \item Complete generation $t+1$ and go to Step ii.
\end{enumerate}

Lastly, we define evaluation criteria:
\begin{enumerate}[i.]
    \item Economic growth, as percentage change in consumption generation over generation.
    \item Frequency of recessions, defined as three consecutive generations of negative growth.
    \item Agent mortality, with the baseline Strategy 0 being the reference 100\%.
    \item Coefficient of variation (CV) of consumption~\cite{champernowne1998economic}, 
    with Strategy 0 being the reference 100\%.
    \item Percentage of failed societies (out of one hundred total).
\end{enumerate}

%==================================
\section{Experiment Results and Discussion}\label{sec:results}
%==================================
Table~\ref{tbl:results} summarizes our experiment results.
Ironically, even though this study was never meant to be an exercise in political economy, the present discussion of results certainly looks like one. First and foremost, the least socially-oriented Strategy~0 has produced societies featuring a modest economic growth of 2.5\% and the best by far level of the society well-being. 
The growth has been very sustainable, with recessions almost non-existent and very few failed societies (i.e. those societies failing to produce another generation of agents before the limit of 100 societies was reached). Population has tended to grow linearly.

\begin{table}[ht]
  \caption{Simulation Results.}
  \label{tbl:results}
  \centering
  \begin{tabular}{l | c | c | c | c}
    \toprule
    \textbf{Metric} & Strategy 0 & Strategy A & Strategy b & Strategy Ab \\ \midrule
    Growth & 2.5\% & 6.3\% & 0.2\% & -2.4\% \\ \midrule
    Recessions & <1\% & 3\% & 12\% & 30\% \\ \midrule
    Mortality & ref. & 156\% & 514\% & 667\% \\ \midrule
    CV & ref. & 114\% & 401\% & 521\% \\ \midrule
    Pct. Failed & 2\% & 5\% & 17\% & 29\% \\
    \bottomrule
  \end{tabular}
\end{table}

Strategy~A, despite its higher level of welfare state\,---\,recall that production gets globally optimized by the ``central planner'' in this strategy\,---\,has shown results reminiscent of Asian Tigers of the past century. Economic growth has certainly been at the Tiger level of 6\%; many societies have experienced population booms. However, the quality of life with Strategy~A is considerably lower than Strategy~0. The biggest contrast is mortality, which has increased by 56\%. The cause is the consistently higher labor input demanded by the more optimal ``central planner''. By construction, a life-long lack of leisure time increases an agent's mortality, to which the 56\% increase is a testimony. Also noteworthy is the 14\% increase in the variability of consumption as well as the higher chances of a recession and a society failure.

\begin{figure}
  \centering
  \includegraphics[width=10cm]{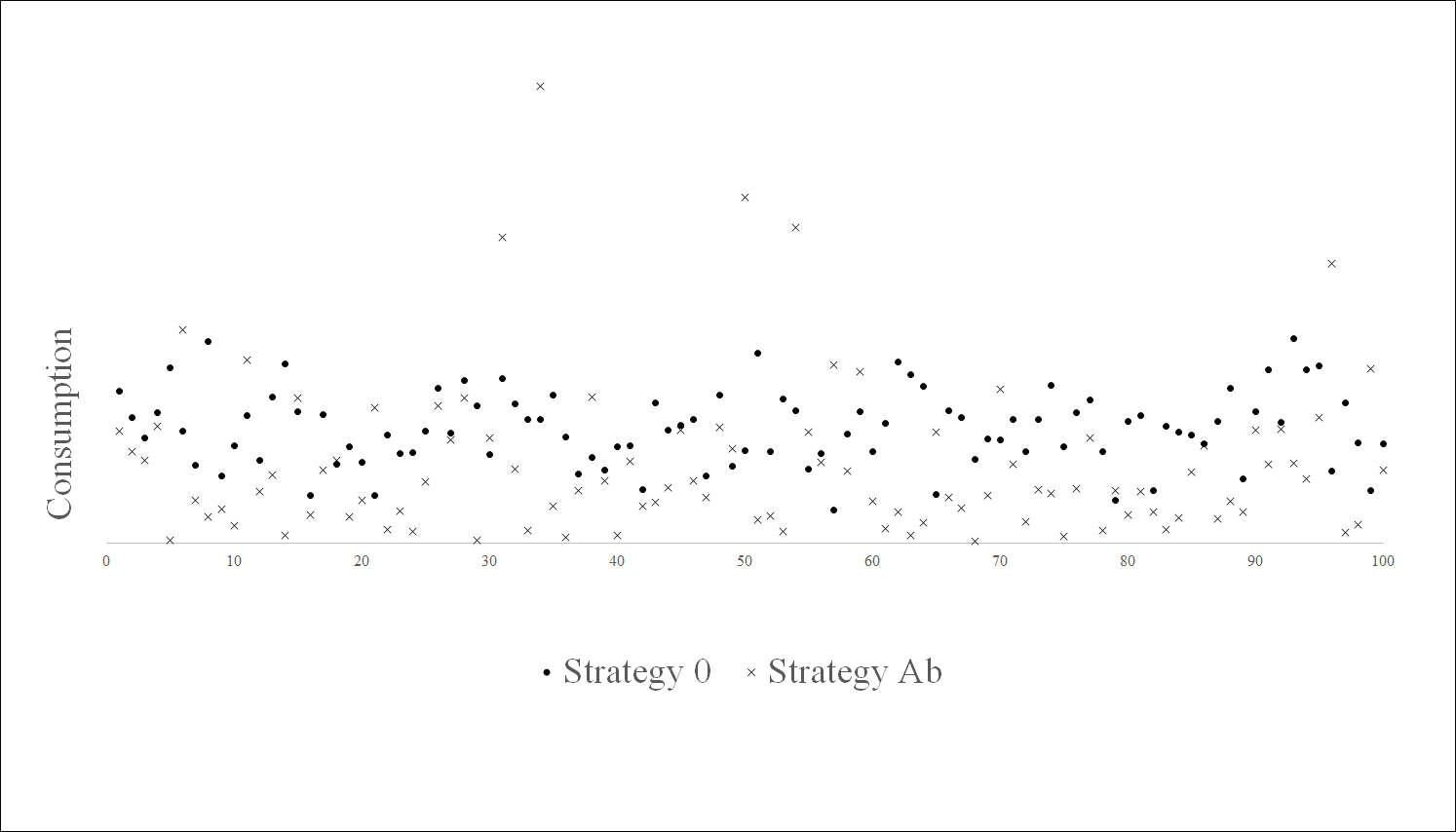}
  \caption{A sample of consumption values of a generation from Strategy~0 versus that from Strategy~Ab. 
  Notice how the Strategy~Ab values are heavily skewed to the right.}\label{fig1}
\end{figure}

In contrast, Strategies~b and~Ab both had a very poor showing. Growth has been nil or negative, agent mortality high, recessions and society failures abundant. Why is this happening given that maximizing the minimum utility in a society seems like a reasonable thing to do, to make the society better? Let us take a closer look at the results of Strategy Ab, the latter being an exponent of the optimization disaster. Figure~\ref{fig1} illustrates the effect of utility maximization with this Strategy. Specifically, a Strategy~Ab society requires an extraordinary amount of labor input to the economy from all of the agents to raise the utility of a relatively few members who have low values of $(\alpha_i,\beta_i,\sigma_i)$; 
in other words, that particular group of individuals does not like anything and is hard to please. Producing unnecessarily high amounts of output drives agent mortality up, 
the surviving agents tend to be mostly from the low utility group, and then the mating of the like plays its role. 
The next generation already is more like the low utility group, and the process thus perpetuates itself. As a result, a population of agents from such a society will likely have a very asymmetric distribution of consumption values: a majority of very low ones and a thick right tail. This makes a stark difference from a Strategy~0 population whose consumption values are tightly scattered around some average value.

%==================================
\section{Concluding Remarks}\label{sec:conclusion}
%==================================
A critic may note\,---\,which will be a fair comment to make\,---\,that the way the experiment was set up may have biased it against the presumably fairer societies, and that the findings are only valid for the specific setup and cannot be generalized in principle. And we wholeheartedly agree, noting in turn that providing a neoliberal expos\'e on socialist planning has never been the, or even a purpose of the study.

Economic and other social interactions are quite complex and diverse to provide a model for any situation or context. Models can help us understand the nature of general trends in society; models can help investigate specific aspects of human behavior and cognition. But in between these extremities lies a wide plateau of problems 
for which there is normally no theory nor empirical data.

As we already alluded to in the introduction, 
ACE-enabled tools are not a mere academic curiosity\,---\,ACE has good practical uses in both business and government. 
Considering building a municipal pool? Designing a new lending facility? 
Upgrading a wireless network or a network of retail locations? 
Test-driving the options in an artificial environment, 
even as simple and basic as one described herein, can provide the much needed quantitative analysis and help uncover potentially costly ``surprises''. 
This is what the present study has aimed to demonstrate.

\bibliographystyle{abbrv}
\bibliography{refs}

\end{document}